\documentclass[prd,superscriptaddress,eqsecnum,nofootinbib,notitlepage,twocolumn,longbibliography,10pt,aps]{revtex4-1}

\usepackage{comment}
\usepackage{enumerate}
\usepackage{mathrsfs}
\usepackage{units}
\usepackage{hyperref,graphicx,amsfonts,amssymb,amsthm,amsmath,psfrag}
\usepackage[toc,page]{appendix}
\usepackage{subfigure}
\usepackage{slashed}
\usepackage{tikz}
\usepackage{booktabs}
\usepackage{siunitx}
\usepackage{bm}       
\usepackage{color}
\usepackage{amssymb}
\usepackage{amsmath}     
\usepackage{hyperref} 
\usepackage{psfrag}
\usepackage[cp1250]{inputenc}
\usepackage{graphicx}
\usepackage{nccmath}
\usepackage[export]{adjustbox}
\DeclareMathAlphabet{\mathpzc}{OT1}{pzc}{m}{it}
\usepackage{booktabs}
\usepackage[first,bottomafter,timestamp]{draftcopy}
\usepackage[normalem]{ulem}	
\newcommand\redout{\bgroup\markoverwith
{\textcolor{red}{\rule[.5ex]{8pt}{0.8pt}}}\ULon}
\usepackage{xcolor}
\usepackage{todonotes}

\usetikzlibrary{decorations.markings,snakes}
\tikzset{
  fermionline/.style={line width=1pt,postaction={decorate},
    decoration={markings,
      mark=at position 0.5 with {\draw[-stealth] (0,0)--(2pt,0);}}},
  bosonline/.style={line width=1pt,decorate,
    decoration={snake,amplitude=1,segment length=4}},
  higgsline/.style={line width=1pt,dashed}
}

\newcommand{\be}{\begin{equation} }
\newcommand{\ee}{\end{equation}}

\newcommand{\nbl}{n_{B-L}}

\begin{document}
\title{Quintessential Affleck-Dine baryogenesis with non-minimal couplings}

\author{Dario Bettoni}
\email{d.bettoni@thphys.uni-heidelberg.de}  
\affiliation{Institut f\"ur Theoretische Physik, Ruprecht-Karls-Universit\"at Heidelberg, \\
Philosophenweg 16, 69120 Heidelberg, Germany}
\author{Javier Rubio}
\email{j.rubio@thphys.uni-heidelberg.de}  
\affiliation{Institut f\"ur Theoretische Physik, Ruprecht-Karls-Universit\"at Heidelberg, \\
Philosophenweg 16, 69120 Heidelberg, Germany}

\begin{abstract}

We present a novel Affleck-Dine scenario for the generation of the observed baryon asymmetry of the Universe based on the non-trivial interplay between quintessential inflationary models containing a kinetic dominated post-inflationary era and a non-minimally coupled $U(1)$ field with a weakly broken $B-L$ symmetry. The non-minimal coupling to gravity renders heavy the Affleck-Dine field during inflation and avoids the generation of isocurvature fluctuations. During the subsequent kinetic era the Ricci scalar changes sign and the effective mass term of the Affleck-Dine field becomes tachyonic. This allows the field to dynamically acquire a large expectation value. The symmetry of the Affleck-Dine potential is automatically restored at the onset of radiation domination, when the Ricci scalar approximately equals zero.  
This inverse phase transition results in the coherent oscillation of the scalar field around the origin of its effective potential. The rotation of the displaced Affleck-Dine field in the complex plane generates a non-zero $B-L$ asymmetry which can be eventually converted into a baryon asymmetry via the usual transfer mechanisms. 
\end{abstract}
\maketitle
\section{Introduction}

One of the open issues in modern particle physics and cosmology is the origin of the observed matter-antimatter asymmetry of the Universe at temperatures below the MeV scale.  This asymmetry is traditionally quantified in terms of the so-called baryon-to-entropy ratio \cite{Ade:2015xua}
\begin{equation}\label{Yb}
Y_B\equiv \frac{n_B}{s} = (0.861\pm 0.008)\times 10^{-10}\,,
\end{equation}
with $n_B\equiv n_{b}-n_{\bar b}$, $n_{b(\bar b)}$ the number density 
of (anti)baryons and $s$ the entropy density. The exponential dilution during inflation of any initial baryon asymmetry calls for a dynamical generation of the ratio $Y_B$ at some point between the end of inflation and big bang nucleosynthesis. Although the Standard Model of particle physics has in principle all the ingredients for generating a baryonic asymmetry during the electroweak phase transition \cite{Sakharov:1967dj}, it cannot produce a sufficiently large baryon-to-entropy ratio~\cite{Bochkarev:1987wf,Kajantie:1995kf,Gavela:1993ts,Huet:1994jb,Gavela:1994dt}. It seems therefore likely that the generation of this quantity is associated to some new physics beyond the Standard Model. Among the different baryogenesis mechanisms proposed in the literature \cite{Dolgov:1991fr,Riotto:1999yt,Dine:2003ax,Morrissey:2012db,DeSimone:2016juo,DeSimone:2016bok}, the Affleck-Dine (AD) scenario \cite{Affleck:1984fy,Dine:1995kz} is particularly interesting, since it can be easily implemented in many beyond the Standard Model extensions involving coherent scalar fields.

The AD mechanism was initially formulated in the context of supersymmetric field theories \cite{Affleck:1984fy,Dine:1995kz}. One of the key ingredients of this proposal is a complex scalar field $\chi$ charged under a weakly broken $U(1)_{B-L}$ symmetry and displaying a sufficiently large (coherent) vacuum expectation value at early times. In the conventional scenario, the initial displacement of the field is generated by a negative Hubble-induced mass term $-c\, H^2\chi^\dagger \chi$, typically associated with supergravity effects and present during \textit{both} the inflationary and radiation-dominated eras. This tachyonic mass term leads to spontaneous symmetry breaking and the associated appearance of a Goldstone mode. If not suppressed, the fluctuations of this massless degree of freedom during inflation might produce isocurvature fluctuations, posing a challenge to the implementation of the Affleck-Dine mechanism within large-field inflationary scenarios \cite{Enqvist:1999hv,Kasuya:2008xp,Kawasaki:2001in}. The expansion of the Universe after inflation leads to the evanescence of the Hubble-induced mass term $-c\, H^2\chi^\dagger \chi$, which  eventually becomes comparable to the \textit{soft}/bare mass of the AD field. When that happens, the radial component of the AD field starts oscillating around the origin of its effective potential and its phase acquires a non-trivial dynamics. The dependence of the $B-L$ charge density on the phase velocity translates into the generation of a net $B-L$ asymmetry, which is eventually converted into a baryon asymmetry. Several variations of the AD mechanism have been proposed in the literature \cite{Sakstein:2017lfm,Sakstein:2017nns,Hertzberg:2013jba,Hertzberg:2013mba}. 

In this paper, we put forward a novel AD scenario 
based on the interplay between non-oscillatory quintessential inflationary models and a non-minimally coupled AD field with a weakly broken $U(1)_{B-L}$ symmetry. In contrast with the standard AD mechanism, the non-minimal coupling to gravity renders heavy the AD field during the inflationary epoch and confines it to the origin of its effective potential. This prevents the generation of sizable isocurvature perturbations. A non-zero expectation value for the AD field is only generated after the end of inflation, where the absence of a potential minimum for the inflaton field leads to a kinetic dominated expansion with negative Ricci curvature. The change of sign of the scalar curvature translates into the appearance of a negative Hubble-induced mass for the non-minimally coupled AD field, which evolves towards a large expectation value.  At the onset of the hot Big Bang era the Ricci scalar vanishes and the non-minimal coupling to gravity effectively disappears. When that happens, the origin of the AD potential becomes again the true minimum and the AD field eventually starts spiraling in the complex plane, creating a non-zero $B-L$ charge density. The subsequent evolution of this quantity and its eventual transfer to the Standard Model particles depends on the particular embedding of the AD field on beyond the Standard Model scenarios. 

This paper is organized as follows. In Section \ref{sec:quint_inf} we review the quintessential inflationary paradigm with special emphasis on the behavior of the Ricci scalar during the different cosmological epochs. The effect of quintessential inflation on the evolution of a non-minimally coupled AD field is discussed in Section \ref{sec:ADmechanism}.  Section \ref{sec:Bgen} contains estimates for the generated $B-L$ charge density as a function of the model parameters. In Section \ref{sec:conclusions} we draw our conclusions. 

\section{Quintessential inflation}\label{sec:quint_inf} 
\noindent Quintessential inflation employs a single degree of freedom to provide a unified description of inflation and dark energy \cite{Peebles:1998qn,Spokoiny:1993kt,Brax:2005uf,Hossain:2014xha,Agarwal:2017wxo,Geng:2017mic,Dimopoulos:2017zvq,Rubio:2017gty}. The simplest realizations of this paradigm make use of a canonical scalar field $\phi$ with Lagrangian density
\begin{equation}\label{cosmonL}
\frac{\cal L}{\sqrt{-g}}=\frac{M_P^2}{2}R -\frac{1}{2}\partial_\mu \phi \partial^\mu \phi -U(\phi)\,,
\end{equation}
where $M_P=(8\pi G)^{-1/2}$ is the reduced Planck mass and $R$ is the Ricci scalar.  We will refer to this inflaton/quintessence field as \textit{cosmon} \cite{Peccei:1987mm}. The cosmon potential $U(\phi)$ is taken to be a non-oscillatory or runaway potential, like the one typically appearing in quintessence and \textit{variable gravity} scenarios \cite{Wetterich:1987fm,Wetterich:1994bg,Wetterich:2013jsa,Wetterich:2014gaa,Hossain:2014xha,Rubio:2017gty}. The scalar field $\phi$ is assumed to be initially displaced at large field values, allowing  for inflation with the usual chaotic initial conditions.  
With a canonical kinetic sector,\footnote{For implementations involving non-canonical kinetic terms see, for instance, Refs.~\cite{Wetterich:2014gaa,Hossain:2014xha,Rubio:2017gty,Dimopoulos:2017zvq,Dimopoulos:2017tud,Akrami:2017cir,Garcia-Garcia:2018hlc}.} a given model within the quintessential inflationary paradigm is specified by a choice of the potential. Several forms of $U(\phi)$, varying from polynomial \cite{Peebles:1998qn} to exponential \cite{Spokoiny:1993kt,Brax:2005uf,Hossain:2014xha,Geng:2017mic,Rubio:2017gty} or hyperbolic potentials \cite{Agarwal:2017wxo}, have been considered in the literature. Although different shapes give rise to different predictions for the inflationary observables and the dark energy equation-of-state parameter, the global evolution of the Universe in these models admits \textit{always} a universal description in terms of the behaviour of the Ricci scalar at the different cosmological epochs.

For a general Friedmann-Lema\^itre-Robertson-Walker background evolution with constant equation-of-state parameter $w$, the Ricci scalar takes the form 
\begin{equation}
R=3(1-3w) H^2\,.
 \end{equation}
During the inflationary stage, the effective equation of state is very close to that associated to a de Sitter expansion, $w= -1$, and $R$ is positive definite,
\begin{equation}\label{Rinf}
R=12 H^2\,.
\end{equation}
Inflation ends when the slow-roll conditions are violated. In the absence of a potential minimum the system enters a \textit{kination} or \textit{deflation} period \cite{Spokoiny:1993kt} in which the total energy budget of the Universe is dominated by the kinetic energy density of the cosmon field. During this regime the effective equation of state approaches $w=1$ and the Ricci scalar changes sign,
\begin{equation}\label{Rkin}
R=-6 H^2\,.
\end{equation}
The kinetic domination regime has to end before big bang nucleosynthesis. In order for this to happen, the energy density of the cosmon must be transmitted to the Standard Model particles in one way or another. Note, however, that, contrary to what happens in oscillatory models of inflation, the decay of the inflaton field does not need to be complete. Even if the energy density of the created particles is initially very small, it will eventually dominate the background evolution due to the rapid decrease of the cosmon energy density during kination domination ($\rho_\phi\sim a^{-6}$) as compared to the radiation component $(\rho_{\rm R}\sim a^{-4})$. 

Several heating mechanisms for quintessential inflationary models have been proposed in the literature \cite{Ford:1986sy,Damour:1995pd,Peebles:1998qn,Felder:1999pv,Feng:2002nb,BuenoSanchez:2007jxm,Rubio:2017gty,Dimopoulos:2018wfg,Nakama:2018gll}.  
The simplest one is based on gravitational particle creation \cite{Spokoiny:1993kt,Ford:1986sy,Damour:1995pd}. Non-conformally coupled scalar fields -- such as the Higgs field, the cosmon or those appearing in beyond the Standard Model extensions such a grand unification -- are inevitably produced in an expanding Universe, provided that they are light enough as compared to the instantaneous Hubble rate. If allowed to interact after production, these scalars lead to a thermal bath and the consequent onset of radiation domination. In spite of its simplicity, this gravitational heating mechanism cannot be considered satisfactory, since it typically requires the existence of a large number of light scalar fields \cite{Rubio:2017gty}, which may lead to dangerous effects such as the generation of large isocurvature perturbations or the  appearance of secondary inflationary periods \cite{Felder:1999pv}. These problems can be avoided by introducing \textit{direct} couplings among the cosmon field and the matter sector.
A highly efficient mechanism within this category is the so-called \textit{instant heating} mechanism \cite{Felder:1999pv}. In this  scenario, the inflaton field $\phi$ is coupled to
some scalar field $X$, which is itself coupled to fermions via Yukawa-type interactions. The growth of the masses of the $X$ particles due to the runaway cosmon field amplifies their energy density and their decay probability into fermions, leading to an almost instantaneous radiation-domination onset. Of course, the ``\textit{instant feeding}'' of the created particles is not a necessary condition and slower heating stages interpolating between gravitational and instant heating scenarios can be easily constructed \cite{Rubio:2017gty}. 

In order to stay agnostic about the precise entropy creation process, we will parametrize the particle production after inflation by a \textit{heating efficiency} parameter \cite{Rubio:2017gty}
\begin{equation}\label{thetadef}
\Theta\equiv \frac{\rho_{\rm R}^{\rm kin}}{\rho^{\rm kin}_\phi}\,,
\end{equation}
with $\rho^{\rm kin}_\phi$ and $\rho_{\rm R}^{\rm kin}$ denoting the energy density of the cosmon and the heating products at the onset of kinetic domination. We assume for simplicity that entropy production takes place instantaneously at that time and neglect any subsequent particle creation. The typical values of $\Theta$ per degree of freedom vary between $\Theta\sim 10^{-27}$ in gravitational heating scenarios\footnote{This number assumes a fiducial Hubble rate $H_{\rm kin}\sim 10^7\,{\rm GeV}$ at the onset of kinetic domination.} \cite{Ford:1986sy,Damour:1995pd,Peebles:1998qn} and $\Theta\sim {\cal O}(1)$ in heating scenarios involving matter fields ~\cite{Felder:1999pv,Rubio:2017gty}.\footnote{Note that the heating efficiency cannot be arbitrarily small. The (integrated) nucleosynthesis constraint on the  gravitational wave density fraction \cite{Maggiore:1999vm} translates into a lower bound \cite{Rubio:2017gty}
\begin{eqnarray}\label{GWbound4}
&&\Theta\gtrsim          10^{-25}\left(\frac{H_{\rm kin}}{10^7\,{\rm GeV}}\right)^2\,. \nonumber 
\end{eqnarray}}  
At the early stages of kinetic domination, the energy density in radiation is still subdominant and does not significantly modify the background evolution. We can take into account the scaling of the different energy components to compute the ratio $\rho_R/\rho_\phi$ at any given time during this period in terms of the heating efficiency $\Theta$,
\begin{equation}\label{XAB}
\frac{\rho_R(a)}{\rho_\phi(a)}=\Theta\left(\frac{a}{a_{\rm kin}}\right)^2\,.
\end{equation}
In particular, we can use Eq.~\eqref{XAB} at the beginning of radiation domination epoch, where  $a=a_{\rm rad}$, $\rho_{\rm R}^{\rm rad}=\rho_\phi^{\rm rad}$ and 
 \begin{equation}\label{XAB2}
\left(\frac{a_{\rm kin}}{a_{\rm rad}}\right)^2=\Theta\,.
\end{equation}
The heating efficiency \eqref{thetadef} can be related to the \textit{radiation temperature} $T_{\rm rad}$ at $a=a_{\rm rad}$. This temperature is defined as
\begin{equation}
T_{\rm rad}\equiv \left(\frac{30\,\rho_{\rm R}^{\rm rad}}{\pi^2 g^{\rm rad}_*}\right)^{1/4}\,,
\end{equation}
with  $g^{\rm rad}_*$ the number of relativistic degrees of freedom at that temperature. 
Taking into account Eq.~\eqref{XAB2}, we can write
\begin{equation}\label{Treh}
T_{\rm rad}
= \left(\frac{30\,\Theta^3 \rho_\phi^{\rm kin}}{\pi^2 g_*^{\rm rad}}\right)^{\frac{1}{4}}
=\Theta^{\frac{1}{2}} \left(\frac{g_*^{\rm kin}}{g_*^{\rm rad}}\right)^{\frac{1}{4}}T_{\rm kin}\,,
\end{equation}
with 
\begin{equation}
T_{\rm kin}\equiv \left(\frac{30\,\rho_{\rm R}^{\rm kin}}{\pi^2 g^{\rm kin}_*}\right)^{1/4}\,,
\end{equation}
the temperature of the created particles at the onset of kinetic domination. Numerically, this corresponds to an energy scale \cite{Rubio:2017gty}
\begin{equation}\label{Tradnum}
\frac{T_{\rm rad}}{10^{12}\, {\rm GeV}}\simeq \, \alpha\, 
 \Theta^{3/4} \left(\frac{H_{\rm kin}}{10^{7} \rm GeV}\right)^{1/2}  \,,
\end{equation}
with $\alpha \equiv 8.65 (g_*^{\rm rad})^{-1/4}$ (cf. Table \ref{table1}). 
Below this temperature, the effective equation of state approaches the radiation value $w\simeq 1/3$ and the Ricci scalar vanishes,
\begin{equation}
R=0  \,.
\end{equation} 
The subsequent background evolution of the Universe proceeds according to the usual hot big bang picture, see for instance Ref.~\cite{Rubio:2017gty}. During the first stages of radiation domination the scalar field $\phi$ \textit{freezes} to a constant field value, which translates into a substantial decrease of its kinetic energy density and the resurgence of its potential counterpart. Once the energy density of the scalar field re-approaches the decreasing energy density of the radiation fluid, the evolution of the system settles down to a scaling solution in which the scalar energy density \textit{tracks} the dominant energy component.\footnote{For a summary of the possible scaling solutions in  generalized scalar-tensor theories, see for instance \cite{Amendola:2018ltt}.} An exit mechanism from this tracking regime, leading to a dark energy dominated era, can be easily implemented in several beyond the Standard Model extensions \cite{Wetterich:2007kr,Amendola:2007yx}. 
\begin{table}
\begin{center}
\begin{tabular}{cccc}    \hline
  $\Theta $ & $(g_*^{\rm rad})^{\frac{1}{4}}T_{\rm rad}$ (GeV) & 
 \\\hline
  $10^{-9}$   & $1.5\times 10^{6}$ & $0<\xi\lesssim 0.33$ \\ 
 $10^{-8}$   & $8.6\times 10^{6}$ & $0<\xi\lesssim 0.41$ \\ 
 $10^{-7}$   & $4.8\times 10^{7}$ & $0<\xi\lesssim 0.52$ \\ 
 $10^{-6}$ & $ 2.7\times 10^{8}$ & $0<\xi\lesssim 0.68 $\\
 $10^{-5}$ & $ 1.5 \times 10^{9}$ & $0<\xi\lesssim 0.94$\\
 $10^{-4}$ & $ 8.6 \times 10^{9}$ & $0<\xi\lesssim 1.41 $\\
\hline
\end{tabular}
\end{center}
 \caption{Typical values of the radiation temperature \eqref{Treh} associated to the production of a single degree of freedom ($g_*^{\rm kin}=1$) at $H_{\rm kin}=
 10^{7}\,{\rm GeV}$ with a heating efficiency $\Theta$. The third column displays the values of the non-minimal coupling $\xi$ leading to a maximum $1\%$ backreaction of the AD field on the background dynamics, cf.~Eq.~\eqref{xicond}.}\label{table1}
\end{table}
\section{Non-minimally coupled Affleck-Dine field}\label{sec:ADmechanism} 
In order to study the interplay between baryogenesis and quintessential inflation we must specify the couplings between the cosmon field $\phi$ and the field(s) associated to some relevant global symmetry, such as the baryon number ($B$) or the baryon number minus the lepton number ($B-L$). Several scenarios involving explicit couplings among these species has been proposed in the literature \cite{DeFelice:2002ir,DeFelice:2004uv,Li:2001st} (see also Ref.~\cite{DeSimone:2016bok}). In this work we take an alternative approach and construct an AD baryogenesis scenario involving only (indirect) gravitational interactions between the cosmon and the AD field. More precisely, we assume the existence of a subdominant  $\mathrm{U}(1)_{B-L}$ field $\chi$ with Lagrangian density
\begin{equation}\label{LBL}
\frac{\mathcal{L}_\chi}{\sqrt{-g}} = -\partial_\mu\chi^\dagger \partial^\mu\chi - V(\chi^\dagger \chi,R)\,,
\end{equation}
on top of the cosmon Lagrangian density \eqref{cosmonL}. The effective potential for this field takes the form
\begin{equation}\label{Veff}
V(\chi^\dagger \chi,R)= \left(A+m_\chi^2\right)\chi^\dagger \chi+ \Delta V\,,
\end{equation}
with $m_\chi^2$ a bare mass parameter and $\Delta V$ some set of $B-L$ symmetric higher-order terms that will play no essential role in our discussion. 
The $A$ term in Eq.~\eqref{Veff} accounts for a non-minimal coupling of the AD field to gravity
\begin{equation} \label{Adef}
A\equiv \xi R=3\xi (1-3 w)H^2\,,
\end{equation}
with $\xi\geq  0$ and $w$ the background equation-of-state parameter. This choice of coupling is motivated by quantum field theory computations in curved spacetime \cite{Birrell:1982ix}. 

The sign of $A$ depends on the the dominant energy component of the Universe. For $w<1/3$ ($A>0$),  we have $A+m^2_\chi>0$ and the minimum of the potential \eqref{Veff} is  located at the origin, $ \vert\chi_{\rm min}\vert=0$.  For $w>1/3$ ($A<0$) and $ A +m_\chi^2 < 0$, the minimum of the potential is located at a non-zero expectation value, $\vert \chi_{\rm min}\vert\neq 0$. The precise value of $\vert \chi_{\rm min}\vert$ depends on the particular set of higher order terms in Eq.~\eqref{Veff}. For a simple choice $\Delta V=\lambda\vert\chi\vert^{n}/\Lambda^{n-4}$ with $\lambda>0$, $n\geq 4$ and cutoff scale $\Lambda$ we have 
 \begin{eqnarray}\label{eq:phimin}
\vert \chi_{\rm min}\vert
&\simeq&
    \left[\frac{2\vert A+ m_\chi^2\vert \Lambda^{n-4}}{n \lambda}\right]^{\frac{1}{n-2}}   \,.
\end{eqnarray}

\noindent The global $\mathrm{U}(1)_{B-L}$ symmetry in Eq.~\eqref{LBL} is associated with a conserved charge density
\begin{equation}\label{eq:charge}
\nbl=J_0\equiv-iq(\chi^\dagger\overset{\leftrightarrow}{\partial_0}\chi)=q \varrho^2\dot{\theta}\,,
\end{equation}
where $q$ is the $B-L$ charge and we have made use of a polar representation $\chi = \tfrac{1}{\sqrt{2}}\varrho \,e^{i \theta}$. Equation \eqref{eq:charge} can be interpreted as the angular momentum of the Affleck-Dine field. Taking the derivative of this equation and using the equation of motion for $\theta$,
\begin{equation}\label{eomcons}
\ddot{\theta}+\left(3H+2\,\partial_t\ln \dot \varrho \right)\dot{\theta}=0\,,
\end{equation}
we can derive the conservation equation
\begin{equation}\label{BLcons}
\frac{1}{a^3}\frac{d}{dt}\left(a^3 n_{B-L}\right)=0\,.
\end{equation}
In order to generate the baryon asymmetry \eqref{Yb} this conservation law must be violated. To this end one has to introduce explicit symmetry-breaking terms on top of the AD potential \eqref{Veff}. Let us assume for concreteness that these operators take a polynomial form of order $n$. Making use of the polar representation of the Affleck-Dine field, these terms can be generically written as
 \begin{equation}\label{eq:Vcorr}
\delta V = -\epsilon_0\, \Lambda^{4-n} \varrho^n F(\theta,\theta_0)  \,,
\end{equation}
with $\epsilon_0$ and $\theta_0$ the modulus and phase of a complex symmetry-breaking parameter $\epsilon\equiv \epsilon_0e^{i\theta_0}$, $\Lambda$ a given cutoff scale and $n\geq 4$. For consistency, we assume $\epsilon_0$ to be small. Note that this condition is technically natural \cite{tHooft:1979rat}. The correction \eqref{eq:Vcorr} translates into an effective potential term for $\theta$ in Eq.~\eqref{eomcons},
\begin{equation}\label{eq:eomcons}
\ddot{\theta}+\left(3H+2\,\partial_t\ln \varrho \right)\dot{\theta}=\epsilon_0\,\Lambda^{4-n} \,\varrho^n F'(\theta,\theta_0)  \,,
\end{equation}
with the prime denoting derivative with respect to $\theta$. As before, Eq.~\eqref{eq:eomcons} can be alternatively written in terms of the $B-L$ charge density \eqref{eq:charge},
\begin{equation}\label{eq:neqn}
\frac{1}{a^3}\frac{d}{dt}\left(a^3 n_{B-L}\right)=q\, \epsilon_0\,\Lambda^{4-n}\varrho^n F'(\theta,\theta_0)  \,.
\end{equation}
This simple differential equation can be formally integrated to obtain the baryon number at a given cosmic time, 
\begin{equation}\label{eq:neqnsol}
a(t)^3 n_{B-L}(t)=q\, \epsilon_0\,\Lambda^{4-n}\int^t dt\, a^3(t)\,\varrho^n F'(\theta,\theta_0) \,.
\end{equation} 
As expected, the generation of a $B-L$ asymmetry is related to the  explicit symmetry-breaking parameter $\epsilon_0$ and approaches zero in the $\epsilon_0\to 0$ limit.

\section{Quintessential baryogenesis}\label{sec:Bgen} 
\begin{figure}
\includegraphics[scale=0.47,center]{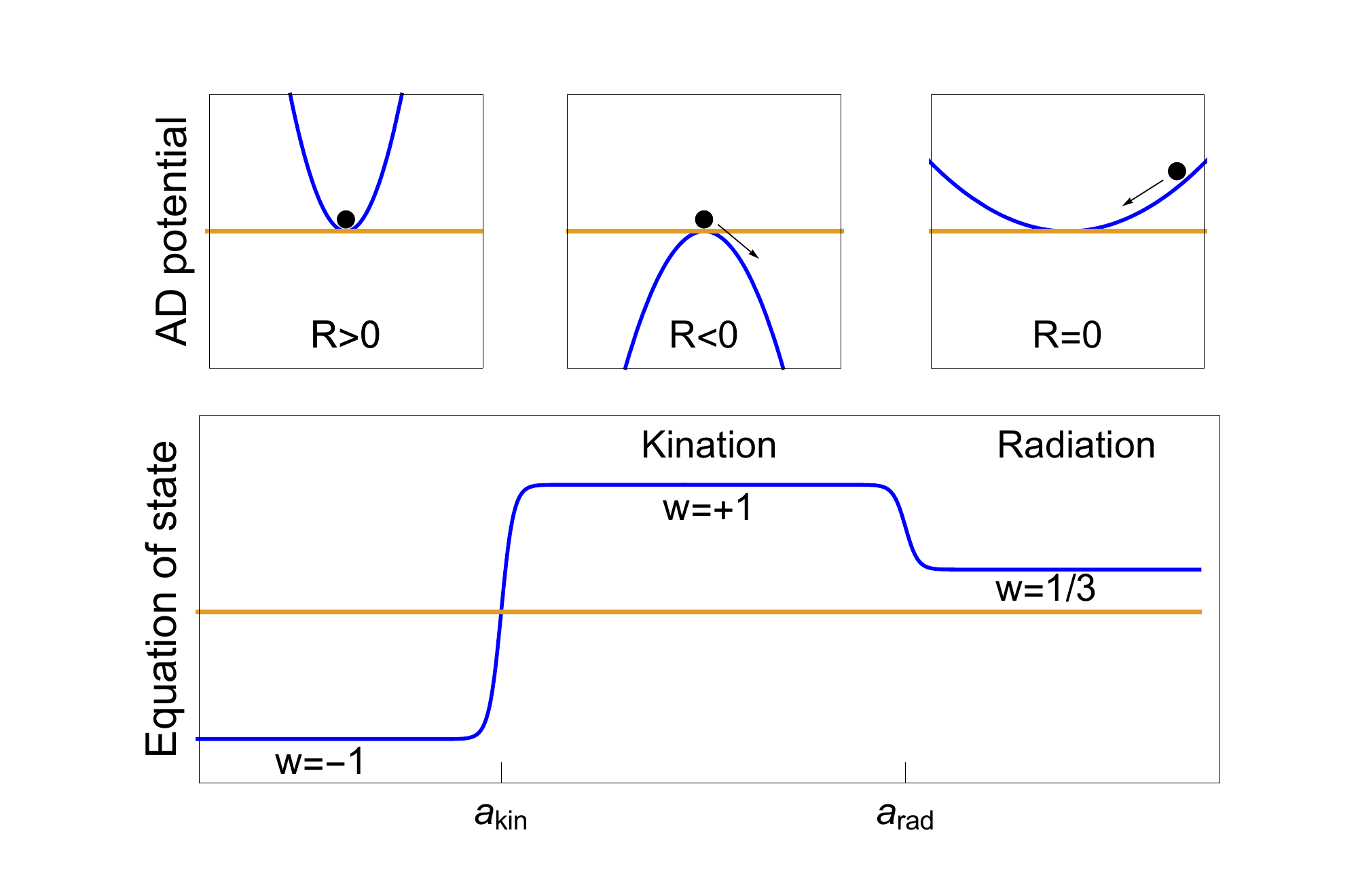}
\caption{Schematic behaviour of the Affleck-Dine potential during the different cosmological epochs in quintessential inflationary scenarios. The horizontal lines in the figure stand for the zero values of the corresponding quantity.} \label{fig:sketch}
\end{figure}

Let us consider the evolution of the non-minimally coupled AD field during the cosmological history presented in Section \ref{sec:quint_inf}. A sketchy illustration of the following discussion is presented in Fig.~\ref{fig:sketch}.

During the inflationary stage ($w=-1$), the $A$ term in Eq.~\eqref{Veff} is positive definite and the AD field is confined to the origin of its effective potential.  In the absence of spontaneous symmetry breaking the two real modes in this complex field are massive. Note that this differs from the usual Affleck-Dine scenario in which the symmetry of the potential is spontaneously broken during inflation. In that case, one of the two components of the Affleck-Dine field becomes a massless Goldstone boson and experiences Brownian motion during inflation, leading to the generation of isocurvature fluctuations and to restrictions on the energy scale of inflation \cite{Kasuya:2008xp}. This limitation is absent in our scenario. In order to avoid the generation of large isocurvature fluctuations we must simply require the effective mass of the AD field to exceed the Hubble rate during inflation. Taking into account Eq. \eqref{Adef}, this translates into a condition
\begin{equation}\label{xicond1}
\xi \gtrsim\frac{1}{\sqrt{12}}\,,
\end{equation}
in the $m_\chi^2\ll A_{\rm inf}$ limit, with $A_{\rm inf}$ the approximately constant value of $A$ during inflation. Under this condition, the $B-L$ charge density \eqref{eq:neqnsol} is either initially zero or rapidly approaches zero due to the exponential expansion of the Universe.

At the onset of the post-inflationary kinetic-dominated era ($w=1$), the effective mass parameter $A$ changes sign. If this tachyonic mass term exceeds the bare mass $m_\chi^2$, the potential \eqref{Veff} develops a hill around $\varrho=0$ and the AD field starts evolving towards large field values. 
The instantaneous expectation value of the $\varrho$ field can be estimated by considering its evolution equation during kinetic domination. Disregarding the bare mass $m_\chi^2$ and the higher-dimensional operators in Eq.~\eqref{Veff}  we get
\begin{equation}
 \quad\ddot{\varrho}+3H\dot{\varrho}+\xi R\varrho=0\quad \rightarrow  \quad\ddot{\varrho}+\frac{1}{t}\dot{\varrho}-\frac{2\xi}{3t^2}\varrho=0\,,
\end{equation} 
where we have made use of Eq.~\eqref{Rkin} with $H=1/(3t)$.
This differential equation admits an exact analytical solution containing a growing and a decaying mode. Taking into account the typical fluctuations of the AD field at the onset on kinetic domination, $\varrho_{\rm kin}=\sqrt{6\xi}H_{\rm kin}/(2\pi)$ and $\dot \varrho_{\rm kin}=H_{\rm kin}\varrho_{\rm kin}$ \cite{Starobinsky:1994bd}, and neglecting the decaying mode, we obtain
\begin{equation}\label{varrhoa}
\varrho(a) \simeq \frac{H_{\rm kin}}{4 \pi} \left(1+\sqrt{6\xi}\right)\left(\frac{a}{a_{\rm kin}}\right)^{\sqrt{6\xi}}\,.
\end{equation}
The growth of the AD field will continue till the moment in which the field finds the minimum 
generated by the higher order terms $\Delta V$ in Eq.~\eqref{Veff} or till the onset of the radiation dominated era, where the Hubble-induced mass term in Eq.~\eqref{Veff} disappears and the symmetry of the AD potential is restored. In what follows we will focus on the second scenario. In this case, the maximum excursion of the AD field is determined by Eq.~\eqref{varrhoa}. Taking into account Eq.~\eqref{XAB2} we can write the value of the AD field at the onset of radiation domination as 
\begin{equation}\label{chirad}
\varrho_{\rm rad}=\frac{H_{\rm kin}}{4 \pi} \left(1+\sqrt{6\xi}\right)\Theta^{-\sqrt{6\xi}/2}\,.
\end{equation} 
In order to ensure a small backreaction effect on the background evolution we require the non-minimal coupling correction to the Planck mass and the energy density of the AD field to stay small, namely\footnote{In this approximation, there is no practical difference between analyzing the problem in the non-minimally coupled frame \eqref{LBL} or in the Einstein frame in which the gravitational part of the action takes the usual Einstein-Hilbert. Indeed, by performing a conformal transformation $g_{\mu\nu}\to \Omega^{-2} g_{\mu\nu}$ with conformal factor $\Omega^2=M_P^2(1-\xi\rho^2/(
M_P^2))$ one obtains an Einstein-frame action coinciding with the original one, up to a set of highly suppressed higher-dimensional operators.}
\begin{equation}\label{noBR}
\xi \varrho^2_{\rm rad}\ll M_P^2\,,\hspace{10mm} \rho^{\rm rad}_{\varrho}\ll \rho^{\rm rad}_{R}\,,
\end{equation}
with $\rho_\varrho^{\rm rad}$ and $\rho_R^{\rm rad}$ the energy densities of the $\varrho$ field and the radiation component at that time. These two conditions translate into a consistency relation involving the non-minimal coupling $\xi$ and the heating efficiency $\Theta$,
\begin{equation}\label{xicond}
\frac{\Theta^{\sqrt{6\xi}/2}}{\sqrt{\xi}+\sqrt{6}\xi}\gg  3\times 10^{-13} \left(\frac{H_{\rm kin}}{10^{7}\,{\rm GeV}}\right) \,,
\end{equation}
which can be easily satisfied for moderate $\xi$ values and not too small heating efficiencies. Typical values of $\xi$ and $\Theta$ leading to a maximum $1\%$ backreaction of the AD field on the background dynamics are shown in Table \ref{table1}.    

If the curvature of the AD potential at symmetry restoration is sufficiently large as compared to the Hubble rate at that epoch ($m_\chi\sim H_{\rm rad}$), the AD field will immediately start oscillating around the origin of its effective potential. In this case, we can derive a consistency relation among the bare mass parameter $m_\chi$, the heating efficiency $\Theta$ and the Hubble rate at the onset of kinetic domination $H_{\rm kin}$,
\begin{equation}\label{mthetarel}
\frac{m_\chi}{{10\,\rm GeV}} \sim \left(\frac{\Theta}{10^{-4}}\right)^{3/2} \left(\frac{H_{\rm kin}}{10^{7}\, {\rm GeV}}\right)\,.
\end{equation}
If, on the contrary, the curvature of the AD potential at symmetry restoration is not sufficiently large ($m_\chi< H_{\rm rad}$), the AD field will stay frozen at its restoration value \eqref{chirad} for 
\begin{equation}\label{DN}
\Delta N=\frac{1}{2}\ln\left(\frac{H_{\rm rad}}{m_\chi}\right)
=\frac{1}{2}\ln\left(\frac{\Theta^{3/2} H_{\rm kin}}{m_\chi}\right)
\end{equation}
e-folds and will start oscillating afterwards. In this case, the consistency relation \eqref{mthetarel} is lost. 

The rotation of the AD field $\chi$ in the complex plane with a large amplitude $\varrho_{\rm rad}$ translates into the generation of a non-zero $B-L$ charge density via Eq.~\eqref{eq:neqn}. In order for the mechanism to be efficient, the right-hand side of Eqs.~\eqref{eq:eomcons} and \eqref{eq:neqn} must be active when the AD field starts oscillating. If the effective mass of the $\theta$ field,
\begin{equation}
 m_\theta \sim \sqrt{\frac{\vert \Delta V'' \vert}{\varrho^2}}\,,
\end{equation}
is smaller than the Hubble rate at that time ($m_\theta<H_{\rm osc}$), the phase will not evolve when the amplitude of $\varrho$ is still large enough to generate a significant $B-L$ asymmetry. We are left therefore with a scenario in which $m_\theta$ is commensurable to $H_{\rm osc}$, namely 
\begin{equation}\label{osc_cond}
H_{\rm osc}\lesssim m_\theta\sim \left(\epsilon_0 \,\Lambda^{4-n}\varrho^{n-2}_{\rm rad}\right)^{1/2}\,,
\end{equation}
with
\begin{equation}\label{Hosccases} 
H_{\rm osc}=
\begin{cases}
\Theta^{3/2} H_{\rm kin}
\hspace{5mm} {\rm if} \hspace{4.5mm}  m_\chi\sim H_{\rm rad}\,,\\
m_{\rm \chi} \hspace{14mm} {\rm if}\hspace{5mm}  m_\chi < H_{\rm rad}\,,
\end{cases}
\end{equation}
This condition translates into a fine-tuning of the dimensionless coupling $\epsilon_0$. 

 The temporal scale for $B-L$ generation is of the order the inverse Hubble parameter at the onset of oscillations, $H^{-1}_{\rm osc}$. Soon after this time the amplitude of $\varrho$ drops down due to the expansion of the Universe and the explicit $B-L$ symmetry-breaking terms in Eqs.~\eqref{eq:eomcons} and \eqref{eq:neqn} become irrelevant, ensuring  the conservation of the created charge density. The short duration of the $B-L$ generation process allows us to obtain an analytic solution for Eq.~\eqref{eq:neqn}. Approximating $\dot{n}_{B-L}\approx H_{\rm osc}\nbl$ \cite{Asaka:2000nb,vonHarling:2012yn}, we get
\begin{equation}\label{nbl}
 \frac{a^3 \nbl }{a^{3}_{\rm osc}} \simeq\frac{q\,\epsilon_0\,\Lambda^{4-n} \varrho_{\rm rad}^n}{ 4 H_{\rm osc}}F'(\theta_{\rm rad},\theta_0)\,.
\end{equation} 
This quantity should be understood as a \textit{local} value of the $B-L$ number density within a given causal patch. Whether a \textit{global} asymmetry is generated or not depends on the precise form of the explicit symmetry-breaking operators. One of the simplest possibilities would be to consider a $Z_4$ symmetric term
\begin{equation}\label{SBterms}
\delta V=\frac{1}{2}\left(\epsilon^*{\chi^\dagger}^4+\epsilon \chi^4\right)\,. 
\end{equation}
For this type of operator we have $F'(\theta,\theta_0)=\sin(4\theta+\theta_0)$ and the 
vacuum manifold does not select a preferred value for $\theta$ prior to symmetry restoration. This conclusion is unchanged if one replaces the $Z_4$ symmetry by a $Z_n$ symmetry or if one takes into account loop corrections. The discrete $Z_{n}$ symmetry forbids the generation of non $Z_{n}$-symmetric operators, such as ($\chi^\dagger\chi^{n-1}+\chi\chi^{\dagger n-1}$). Only $Z_n$-symmetric higher-order operators such as $\lambda^2(\chi^{\dagger 2n}+\chi^{2n})/(M^2_P+2\xi\chi^\dagger\chi)^2$ can be generated by loop effects \cite{Bezrukov:2012hx}. In this case, even if an asymmetry can be \textit{locally} generated,  one should expect an almost zero \textit{global} asymmetry and therefore no successful baryogenesis.\footnote{Note, however, that this case can give rise to an interesting phenomenology involving the generation of a cosmic string network and the associated production of gravitational waves \cite{Kamada:2015iga}. When heating takes place, the AD field starts oscillating around the origin and topological defects are expected to disappear.} We emphasize that this is not a particular limitation of our scenario, but rather a generic feature of any post-inflationary Affleck--Dine mechanism involving a spontaneous symmetry-breaking potential, as that considered for instance in Refs.~\cite{Sakstein:2017lfm,Sakstein:2017nns}. 

In order to generate a non-vanishing global asymmetry one could consider, for instance, non $Z_n$-symmetric operators with odd $n$ leading to a preferred value for the $\theta$ field. Since $n_{B-L}\propto \dot \theta$ (cf. Eq.~\eqref{eq:charge}), it is also necessary that the number of clockwise configurations when all causal patches are taken into account  exceeds the number of anticlockwise configurations (or the other way around). For this  to happen, the ground state of the system prior to symmetry restoration should be approached in an asymmetric way. This can be achieved by a combination of non $Z_n$-symmetric operators with complex coefficients $\epsilon_n$. If the mass of the $\theta$ field slightly exceeds the Hubble rate before symmetry restoration, the AD field  will explore the instantaneous minima associated to the dominant symmetry-breaking operator at a given $\varrho$. If the complex coefficients $\epsilon_n$ are different, the instantaneous minima in the $\theta$ direction will not be  perfectly aligned for all values of $\varrho$. In that case, the dynamics of the phase will be dominated by a classical driving force. This will generate a non-zero ``bulk'' velocity for the $\theta$ field on top of its stochastic motion, such that when the symmetry is restored the number of clock- and anticlock-wise configurations will be generically different and a non-vanishing \textit{global} asymmetry can be generated. Determining the precise phase dynamics would require the specification of the particular set of non $Z_{n}$-symmetric operators and probably the use of numerical simulations. We postpone this detailed analysis to a future work and provide here just an order of magnitude estimate for the \textit{global} $B-L$ asymmetry. To this end, we will assume a generic value $F'(\theta_{\rm rad},\theta_0)\sim f$, with $f$ representing the level of asymmetry in the phase distribution prior to symmetry restoration. Under this assumption Eq.~\eqref{nbl} can be understood as an equation for the \textit{global} $B-L$ asymmetry. Assuming $m_\theta\sim {\cal O}(H_{\rm osc})$, we obtain the estimate
\begin{equation}\label{nblcases}
\nbl   \simeq \frac{f\, q\, H_{\rm osc}^3}{4\epsilon_0^{\beta}} 
\left(\frac{\Lambda}{H_{\rm osc}}\right)^{\gamma} \left(\frac{a_{\rm osc}}{a}\right)^3\,, 
\end{equation}
with $H_{\rm osc}$ given by Eq.~\eqref{Hosccases}, $\beta\equiv 2/(n-2)$ and $\gamma\equiv (n-4)\beta$. Dividing the $B-L$ charge density \eqref{nblcases} by the entropy density of the radiation component.
\begin{equation}
s=\frac{4\rho_{\rm R}}{3T_{\rm R}}=\frac{4 M_P^2 H^2_{\rm rad}}{T_{\rm rad}}\left(\frac{a_{\rm rad}}{a}\right)^3 \,,
\end{equation}
we can define the dimensionless ratio
\begin{equation}\label{Ydef}
Y_{B-L}\equiv \frac{n_{B-L}}{s}\,.
\end{equation} 
Note that since both $n_{B-L}$ and $s$ decrease as $a^{-3}$ as the universe expands, $Y_{B-L}$ takes a constant value as long as no further entropy production takes place. At this point we can consider two scenarios, associated to the two cases in Eq.~\eqref{Hosccases}. If the bare mass of the AD field is commensurable with the Hubble rate at the onset of radiation domination, $m_\chi \sim H_{\rm rad}$, we have $a_{\rm osc}\simeq a_{\rm rad}$ and $H_{\rm osc}\simeq H_{\rm rad}$. In this case the dimensionless ratio $Y_{B-L}$ can be written as 
\begin{equation}\label{YBL0}
Y_{B-L}\simeq \frac{1}{10}
\frac{f\,q}{g_*^{1/4}\epsilon_0^\beta} \left(\frac{H_{\rm rad}}{M_P}\right)^{\frac{3}{2}} \left(\frac{\Lambda}{H_{\rm rad}}\right)^{\gamma} \,.
\end{equation}
If instead the mass of the AD field is smaller than the Hubble rate at the onset of radiation domination, $m_\chi < H_{\rm rad}$, the dimensionless ratio $Y_{B-L}$ becomes
\begin{equation}\label{YBL2}
Y_{B-L}\simeq \frac{1}{10}
\frac{f\,q}{g_*^{1/4}\epsilon_0^\beta} \left(\frac{H_{\rm rad}}{M_P}\right)^{\frac{3}{2}} \left(\frac{\Lambda}{m_{\chi}}\right)^{\gamma} \left(\frac{m_{\chi}}{H_{\rm rad}}\right)^{3/2} \,,
\end{equation}
where we have made use of Eq.~\eqref{DN}. In both cases, a small ratio $Y_{B-L}\sim 10^{-10}$ can be easily obtained for a sensible choice of parameters. For instance, for the lowest order $n=4$ in Eq.~\eqref{YBL0} we can choose 
\begin{equation}\label{YBL1}
\frac{Y_{B-L}}{10^{-10}} \simeq 
\frac{f\,q}{g_*^{1/4}} \left(\frac{10^{-17}}{\epsilon_0}\right)\left(\frac{H_{\rm rad}}{10\,{\rm GeV}}\right)^{\frac{3}{2}}\,,
\end{equation}
with  
\begin{equation}
\frac{H_{\rm rad}}{10\,{\rm GeV}}=\left(\frac{\Theta}{10^{-4}}\right)^{\frac{3}{2}}\frac{H_{\rm kin}}{10^{7}\,{\rm GeV}}\,. 
\end{equation} 
Taking into account these prototypical values we can estimate the displacement $\varrho_{\rm rad}$ and the non-minimal coupling needed to achieve it before symmetry restoration. Using Eqs.~\eqref{chirad} and \eqref{osc_cond} we obtain a sub-Planckian field value $\varrho_{\rm rad}\sim 10^{-9} M_P$ and a non-minimal coupling $\xi \sim {\cal O}(1)$, in good agreement with the consistency relations \eqref{xicond1} and \eqref{xicond}.

In order to generate a baryonic asymmetry, the created $B-L$ asymmetry must be first transferred to the left-handed Standard Model particles. The generation of this primordial leptonic asymmetry  can take place through a variety of processes that depend on the particular embedding of the AD field on beyond the Standard Model scenarios. A potential transferring mechanism involving the neutrino portal operator \cite{Falkowski:2009yz,Macias:2015cna} was identified in Ref.~\cite{Sakstein:2017lfm}. Once the leptonic asymmetry is produced, the sphaleron effects operating above the electroweak scale will reprocess and convert it into a non-zero baryonic number \cite{Kuzmin:1985mm,Fukugita:1986hr,Harvey:1990qw}. 

\section{Conclusions}\label{sec:conclusions}
Quintessential inflation is a rather minimalistic paradigm involving a single degree of freedom, the cosmon, for explaining both the inflationary epoch and the dark energy dominated era.  In this work, we have put forward a novel way of generating the baryon asymmetry of the Universe in this type of models. Contrary to other proposals in the literature, our mechanism does not make use of any direct coupling among the cosmon and the field(s) carrying baryon-lepton number. In the spirit of the original Affleck-Dine mechanism, we only consider indirect couplings among these species. In particular, we assumed the existence of a subdominant non-minimally coupled scalar field with a weakly broken $U(1)_{B-L}$ symmetry. This Affleck-Dine field couples then to the Ricci scalar evolution determined by the dominant  cosmon energy density but not to the field itself.  The non-minimal coupling to gravity gives rise to a natural realization of Affleck-Dine baryogenesis which displays, however,  some important differences with the standard scenario. On the one hand,  it renders heavy the Affleck-Dine field during inflation and prevents the generation of large isocurvature perturbations. On the other hand, the displacement of the Affleck-Dine  field takes place only after the end of inflation, where the change of sign of the Ricci scalar leads to spontaneous symmetry breaking and the consequent evolution of the  Affleck-Dine field towards non-zero values. The evanescence of the non-minimal coupling at the heating stage gives rise to symmetry restoration and the subsequent oscillation of the  Affleck-Dine field around the origin of its effective potential. This generates a net $B-L$ asymmetry, which can be eventually converted into a baryon asymmetry via standard mechanisms. 

We emphasize that the framework presented in this paper is not restricted to canonically normalized inflationary models with runaway potentials,  but applies to any cosmological scenario involving a post-inflationary era with a \textit{stiff} equation-of-state parameter $1/3 < w\leq 1$, irrespective of its origin. Particular realizations of our idea can take place, for instance, in quantum gravity inspired models \cite{Wetterich:2014gaa,Hossain:2014xha,Rubio:2017gty}, $\alpha$-attractor frameworks \cite{Dimopoulos:2017zvq,Dimopoulos:2017tud,Akrami:2017cir,Garcia-Garcia:2018hlc} or braneworld scenarios  \cite{Copeland:2000hn,Liddle:2003zw}.

\section*{Acknowledgments}

We acknowledge support from DFG through the project TRR33 ``The Dark Universe''. We thank Christof Wetterich and Takeshi Kobayashi for useful comments and discussions. 
\bibliography{QBNMC}

\end{document}